\documentclass[conference]{IEEEtran}
\IEEEoverridecommandlockouts

\usepackage{cite}
\usepackage{amsmath,amssymb,amsfonts}
\usepackage{algorithmic}
\usepackage{graphicx}
\usepackage{textcomp}
\usepackage{multirow} 
\usepackage{xcolor}
\def\BibTeX{{\rm B\kern-.05em{\sc i\kern-.025em b}\kern-.08em
    T\kern-.1667em\lower.7ex\hbox{E}\kern-.125emX}}
\begin{document}
\newcommand{\redacted}[1]{[REDACTED]\footnote[1]{Paper under submission and details omitted for double-blind review.}}
% \title{Assessing Data Augmentation-Induced Bias in Training of Large Language Models (LLMs)\thanks{Funded by X.}
% }
\title{Assessing Data Augmentation-Induced Bias in Training and Testing of Machine Learning Models\thanks{This research was supported by the Natural Sciences and Engineering Research Council of Canada (NSERC), grant 2018-06588, and by MITACS.}
}

\author{\IEEEauthorblockN{Riddhi More}
 \IEEEauthorblockA{\textit{Ontario Tech University}\\
Oshawa, ON, Canada\\
riddhi.more1@ontariotechu.net}
\and
\IEEEauthorblockN{Jeremy S. Bradbury}
\IEEEauthorblockA{\textit{Ontario Tech University}\\
Oshawa, ON, Canada\\
jeremy.bradbury@ontariotechu.ca}

}

\maketitle

\begin{abstract}
% Data augmentation has become a standard practice in AI, including with LLMs where the training data needs can be greater than the available data or where the available data lacks diversity. Data augmentation, including SMOTE and mutation-based augmentation, is already used in LLM-based software testing and debugging applications without a rigorous understanding of the impact of the augmented training data set on model bias and possible overfitting. We propose an empirical approach to assessing augmented training data and demonstrate our approach by assessing the impact of augmented training data in an existing data set for classifying flaky tests.

Data augmentation has become a standard practice in software engineering to address limited or imbalanced data sets, particularly in specialized domains like test classification and bug detection where data can be scarce. Although techniques such as SMOTE and mutation-based augmentation are widely used in software testing and debugging applications, a rigorous understanding of how augmented training data impacts model bias %and potential overfitting 
is lacking. It is especially critical to consider bias in scenarios where augmented data sets are used not just in training but also in testing models. Through a comprehensive case study of flaky test classification, we demonstrate how to test for bias and understand the impact that the inclusion of augmented samples in testing sets can have on model evaluation.
\end{abstract}

\begin{IEEEkeywords} 

data augmentation, data validation, 
machine learning, 
software testing,
mutation testing,
flaky tests,
bias\end{IEEEkeywords}
\section{Introduction}

Data augmentation \emph{``...involves enhancing the sufficiency and diversity of training examples without explicitly collecting new data... The essence of data augmentation lies in generating new data by altering existing data points through various transformations"}~\cite{ZGW+24}. Data augmentation has become a vital technique in software engineering (SE) to address training data limitations (e.g., small data set size, data set imbalance) in machine learning applications. This is particularly true with the advent of large language models in application areas such as code generation~\cite{DQZ+24}. Augmentation methods for software engineering data sets include Synthetic Minority Oversampling Technique (SMOTE)~\cite{Chawla2002}, mutation analysis, % Early research established systematic approaches for data set augmentation in SE using 
neural network-based methods~\cite{JGR19} and more recently the use of LLMs to augment or train data~\cite{YZZ+23,DQZ+24}.

However, a critical methodological concern arises when augmented samples appear in both training and testing sets, potentially inflating performance metrics through pattern recognition rather than true generalization. This issue is especially pronounced in code-based tasks, where augmentation must preserve both syntactic and semantic validity~\cite{YZZ+23}. We present a systematic investigation into augmentation-induced bias in SE models, using flaky test classification as a case study. Through carefully designed experiments, we:

\begin{itemize}
    \item Quantify the extent of performance inflation when augmented samples appear in both training and testing sets
    \item Analyze how different types of code augmentation affect model bias across various categories of flaky tests
    \item Propose guidelines for proper data set splitting when using augmented data to ensure more reliable model evaluation
    \item Demonstrate how controlling for augmentation bias can provide more accurate assessments of model generalization capabilities
\end{itemize}

Our findings reveal differences in model performance when evaluating truly independent test cases versus augmented variants of training samples. These results have important implications for how we assess and report the effectiveness of models trained on augmented data in SE tasks. Furthermore, our analysis provides insight on the development of more robust evaluation methodologies that can better predict real-world model performance.

The rest of this paper is organized as follows: Section II provides background on data augmentation in SE and current evaluation practices. Section III details our experimental methodology for assessing augmentation bias through a case study, findings, and discussion. Section IV concludes with a summary of our contributions and their significance to the SE community. The code is available in our GitHub repository.\footnote{https://github.com/seer-lab/AugmentationBias}

\section{Background}

%\subsection{Data Augmentation in Software Engineering}
%Data augmentation has become an increasingly important technique in software engineering (SE), particularly as machine learning approaches become more prevalent in SE tasks. The technique involves creating synthetic data samples through various transformations of existing data, helping to expand limited data sets and improve model generalization. 

\subsection{Limited and Imbalanced data sets in SE}

The use of machine learning and LLMs to solve specialized software engineering tasks in niche domains is often challenged by limited or imbalanced data sets that result from a lack of available data. %, 
%These problems are prime candidates for data augmentation. 
For example, this challenge is particularly evident in software testing, where the collection of comprehensive test suites covering all possible scenarios is often impractical~\cite{HWRC24}. Recent techniques, including the combination of fuzzing with LLM, have attempted to address the limited diversity in test data sets~\cite{HWRC24}.%, highlighting the persistent challenge of data scarcity in testing scenarios.

Within software testing, the problem of flaky test classification exemplifies these data limitations. Flaky tests are tests that exhibit non-deterministic behavior, often passing and failing inconsistently when run on the same code version\cite{Lam2019}. Research has successfully addressed the scarcity of labeled flaky test examples through augmentation, utilizing techniques like the SMOTE for test code.

In the broader context of software engineering, similar challenges appear in vulnerability assessment, where studies have investigated data augmentation as a potential solution to inherent data set imbalances~\cite{LA24}. These investigations emphasize the critical need to evaluate augmentation techniques' effectiveness in highly specialized SE tasks.

\subsection{Current Practices and Potential Bias}

While data augmentation has been utilized successfully across software engineering, current practices in evaluating models trained on augmented data raise legitimate concerns.% important methodological concerns. 
Comprehensive surveys of large model-based data augmentation have highlighted how these approaches, while outperforming traditional methods, introduce new challenges with respect to the quality and reliability of the augmented data~\cite{ZGW+24}.

A significant concern lies in the common practice of including augmented samples as well as samples used in the generation of augmented samples, in both training and testing data sets. This practice can potentially lead to artificially inflated performance metrics that are not representative of data that is unconnected to augmentation. This practice becomes particularly problematic when variants of the same original sample appear in both the training and testing data sets. In this instance, models may learn to recognize patterns introduced by the augmentation process rather than truly generalizing to solve the underlying task.%new examples.

Despite the widespread use of data augmentation in SE tasks, the potential bias introduced by these practices has not been systematically studied. %The risk is especially pronounced in code-based tasks where subtle variations from augmentation might create recognizable patterns that do not reflect real-world variations. 
This gap in understanding motivates our current case study, which aims to quantify and characterize the impact of augmentation-induced bias in one particular instance of model evaluation. 
% \section{Background}

% \begin{itemize}
%     \item data augmentation in SE
%     \item testing~\cite{HWRC24}
%     \item flaky test classifcation~\cite{AHH+23}
%     \item bug localization~\cite{CD23}
% \end{itemize}

\section{Case Study: Assessing Data Augmentation Bias in FlakyCat Training}

In this case study, we examine the introduction of bias through data augmentation in the context of flaky test classification. We utilize the FlakyCat data set~\cite{AHH+23} and conduct two experiments to assess both the effectiveness and potential drawbacks of augmentation on the classification of flaky tests.

\subsection{data set Composition and Preparation}

To avoid self-confirmatory bias in our case study, we rely on an existing augmentation technique that was applied to the FlakyCat data set by a third-party~\cite{AHH+23}. The FlakyCat dataset consists of five primary categories of flaky tests: async wait (Async), test order dependency (TOD), time, concurrency (Conc), and unordered collections (UC). Each sample includes an original version (v0) and two augmented versions (v1, v2). An adapted Synthetic Minority Over-sampling Technique (SMOTE) was utilized by Akli et al.~\cite{AHH+23} to create the augmented versions. SMOTE is a well-established data augmentation method known for addressing class imbalances effectively. Akli et al. adapted SMOTE for test code by generating variants through controlled mutations of non-flakiness-related elements, such as variable names, constants (e.g., strings), and test method names, as well as adding unused variable declarations. These elements were replaced with randomly generated words to differentiate the augmented version source code from the original source code. 

The adapted SMOTE method used by Akli et al. expanded the FlakyCat data set threefold across all flaky test categories while preserving the original flakiness distribution.%By using a technique developed independently of our study, we ensured fairness and avoided artificial results, focusing instead on evaluating the impact of bias introduced by augmentation.

\subsection{Training Configuration}

For our experiments, we used a model from our previous work on FlakyXBert~\cite{MB25}, which included a Siamese network classifier and CodeBERT~\cite{Guo2020CodeBERT} as its base encoder to generate the initial Flaky test code representations. The model was trained for 200 epochs using the contrastive loss function to distinguish between different root causes of flakiness by comparing pairs of test cases.

The training hyperparameters were:
\begin{itemize}
   \item Learning rate: $1 \times 10^{-5}$
   \item Batch size: 8
   \item Optimizer: Adam with weight decay
\end{itemize}

These hyperparameters were selected to enhance the model's ability to learn from a small number of training examples without overfitting while optimizing the balance between training speed and memory usage.

All experiments were performed on a Dell RTX 4090 workstation with Nvidia drivers (ver.~555) and CUDA~12.5.

The experiments followed two distinct protocols (Experiment 1 and 2 as described in the following sections), with each maintaining its own training and testing splits according to their respective objectives.

%We designed two experiments to systematically evaluate the impact of data augmentation on model bias and performance. Each experiment addresses different aspects of potential augmentation-induced bias.

\subsection{Experiment 1 -- Augmentation Impact Assessment}

\subsubsection{Experimental Design}

% Our experiment aims to systematically evaluate the impact of data augmentation through a two-phase approach using the same initial random sampling of test cases. In both phases, we begin by randomly splitting the data set into training and testing sets using only original test cases. For the baseline assessment (Phase A), we train and evaluate the model using only these original test cases to establish the model's base performance.

% For the augmentation assessment (Phase B), we follow a strict protocol: when an original test case is selected for training, both its augmented versions are also included in the training set. Similarly, if a test case is selected for testing, both its original and augmented versions appear in the test set. The goal is to get a more honest evaluation of the model's ability to generalize to truly unseen test cases and assess whether the model is learning legitimate patterns that identify flaky tests rather than just memorizing artifacts of the augmentation process.

% Through this experimental design, we can directly compare the model's performance with and without augmentation. This comparison allows us to quantify the benefits of augmentation. Table~\ref{tab:exp1} shows the data distribution in the training set and the testing set for Phase B of Experiment 1.

Our experiment evaluates the impact of data augmentation using a two-phase approach with the same initial random sampling of test cases. In both phases, we split the dataset into training and testing sets using only original test cases. In Phase A, we establish the model's baseline performance by training and evaluating it on these original cases.

In Phase B, we follow a strict protocol: when an original test case is selected for training, both of its augmented versions are also included in the training set. Similarly, if a test case is selected for testing, both its original and augmented versions appear in the test set. %The goal is to get a more honest evaluation of the model's ability to generalize to truly unseen test cases and assess whether the model is learning legitimate patterns that identify flaky tests rather than just memorizing artifacts of the augmentation process.
%augmented versions of test cases are included alongside the originals, maintaining consistency between training and testing sets. 
This ensures a more accurate evaluation of the model's ability to generalize to unseen cases and assess whether it learns meaningful patterns rather than artifacts of augmentation.

This design enables a direct performance comparison with and without augmentation, quantifying its benefits. Table~\ref{tab:exp1} details the data distribution for Phase B of Experiment 1.
\begin{table}[h!]
\centering
% \resizebox{\textwidth}{!}{%
\begin{tabular}{l|cccc|cccc}
\hline
\multirow{2}{*}{\textbf{Category}} & \multicolumn{4}{c|}{\textbf{Training Set}} & \multicolumn{4}{c}{\textbf{Testing Set}} \\ \cline{2-9} 
                                   & \textbf{Total} & \textbf{v0} & \textbf{v1} & \textbf{v2} & \textbf{Total} & \textbf{v0} & \textbf{v1} & \textbf{v2} \\ \hline
Async                        & 298            & 102          & 98          & 98        & 78             & 26          & 26          & 26          \\ \hline
UC              & 132            & 44          & 44          & 44          & 21             & 7           & 7           & 7           \\ \hline
Conc                        & 117            & 39          & 39          & 39          & 27             & 9           & 9           & 9           \\ \hline
Time                               & 100            & 34          & 33          & 33          & 22             & 8           & 7           & 7           \\ \hline
TOD              & 221            & 79          & 71          & 71          & 70             & 24          & 23          & 23          \\ \hline

\end{tabular}%
% }
\vspace{1mm}
\caption{Combined File Counts for Training and Testing Sets in Experiment~1}
\label{tab:exp1}
\end{table}

\subsubsection{Results and Analysis}

%Our experiments yielded quantitative insights into the impact of data augmentation on model performance across different flaky test categories. We present these results through F1 scores, which provide a balanced measure of precision and recall.

\begin{figure}[h!]
    \centering
    \includegraphics[width=\linewidth]{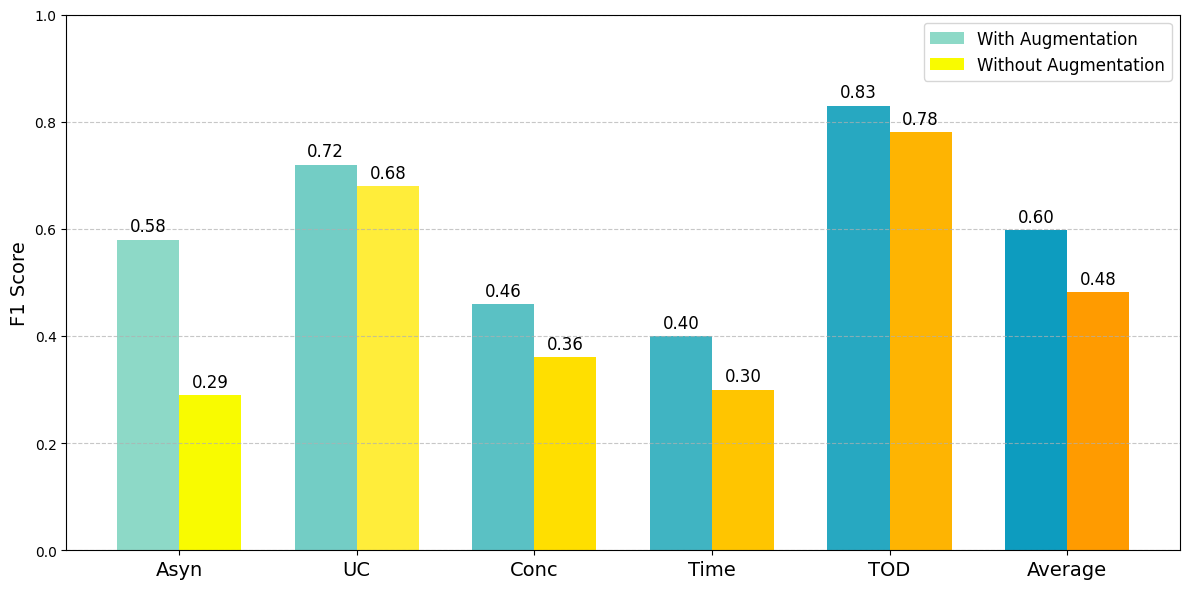}
    \caption{F1 score by category Phase~A vs Phase~B from Experiment~1}
    \label{fig:aug-comparison}
\end{figure}

This experiment evaluated whether data augmentation introduces bias by ensuring that related test cases (original and augmented versions) were kept together in training or testing sets. The results shown in Figure~\ref{fig:aug-comparison} reveal several key insights regarding augmentation's impact.

The model trained with augmented data (Phase B) consistently outperformed the baseline model trained only on original data (Phase A) across all categories, with an average F1 score improvement of 12\% (from 48\% to 60\%). This improvement, achieved while maintaining the integrity of test case relationships, suggests that the augmentation process has an impact on the performance of the model.

Most notably, categories with typically challenging patterns such as Async Wait and Time showed substantial improvements (0.29 to 0.58 and 0.30 to 0.40 respectively), indicating that augmentation helps capture complex patterns. This suggests that augmentation potentially enhances the key characteristics that make these tests flaky.

\subsection{Experiment 2 -- Augmentation Bias Analysis}

\subsubsection{Experimental Design}

% The second experiment evaluates the model's generalization capabilities by distinguishing between its performance on augmented versions of training data and entirely new test cases. We structure the experiment by training the model exclusively on original versions (v0) of test cases, then evaluate its performance on two distinct test sets. The first test set (Testing Set 1) contains only original versions (v0) of previously unseen test cases, while the second test set (testing Set 2) comprises augmented versions (v1 and v2) of the test cases used in training. This experimental design allows us to assess whether the model exhibits different performance characteristics when dealing with augmented variants of familiar test cases versus completely new test cases, thereby providing insights into whether the model truly learns generalizable characteristics of flaky tests or potentially overfits to patterns introduced by the augmentation process. the distribution is shown in Table~\ref{tab:exp2}.

The second experiment assesses the model's generalization by comparing its performance on augmented training data with entirely new test cases. The model is trained exclusively on original test case versions (v0) and evaluated on two test sets: Testing Set 1, containing original versions (v0) of unseen test cases, and Testing Set 2, comprising augmented versions (v1 and v2) of training cases.

This design examines whether the model performs differently on augmented variants of familiar cases versus new cases, providing insight into its ability to generalize flaky test characteristics or its tendency to overfit augmentation patterns. The distribution is detailed in Table~\ref{tab:exp2}.

\begin{table}[h!]
\centering
% \resizebox{\textwidth}{!}{%
\begin{tabular}{l|cc|cc|ccc}
\hline
\multirow{2}{*}{\textbf{Category}} & \multicolumn{2}{c|}{\textbf{Training Set}} & \multicolumn{2}{c|}{\textbf{Testing Set 1}} & \multicolumn{3}{c}{\textbf{Testing Set 2}} \\ \cline{2-8} 
                                   & \textbf{Total} & \textbf{v0} & \textbf{Total} & \textbf{v0} & \textbf{Total} & \textbf{v1} & \textbf{v2} \\ \hline
Async                         & 99             & 99          & 26             & 26          & 251            & 124         & 127         \\ \hline
UC         & 44             & 44          & 7              & 7           & 102            & 51          & 51          \\ \hline
Conc                & 39             & 39          & 9              & 9           & 96             & 48          & 48          \\ \hline
Time                               & 34             & 34          & 8              & 8           & 80             & 40          & 40          \\ \hline
TOD       & 79             & 79          & 24             & 24          & 188            & 94          & 94          \\ \hline

\end{tabular}%
% }
\vspace{1mm}
\caption{Combined File Counts from Training and Testing Sets in Experiment~2 (with zero-only columns removed).}
\label{tab:exp2}
\end{table}

\subsubsection{Results and Analysis}

This experiment evaluated bias in the augmentation process by comparing model performance on augmented training data versus entirely new test cases, focusing on whether the model learns generalizable characteristics of flaky tests or becomes biased by patterns introduced through augmentation.

\begin{figure}[h!]
    \centering
    \includegraphics[width=\linewidth]{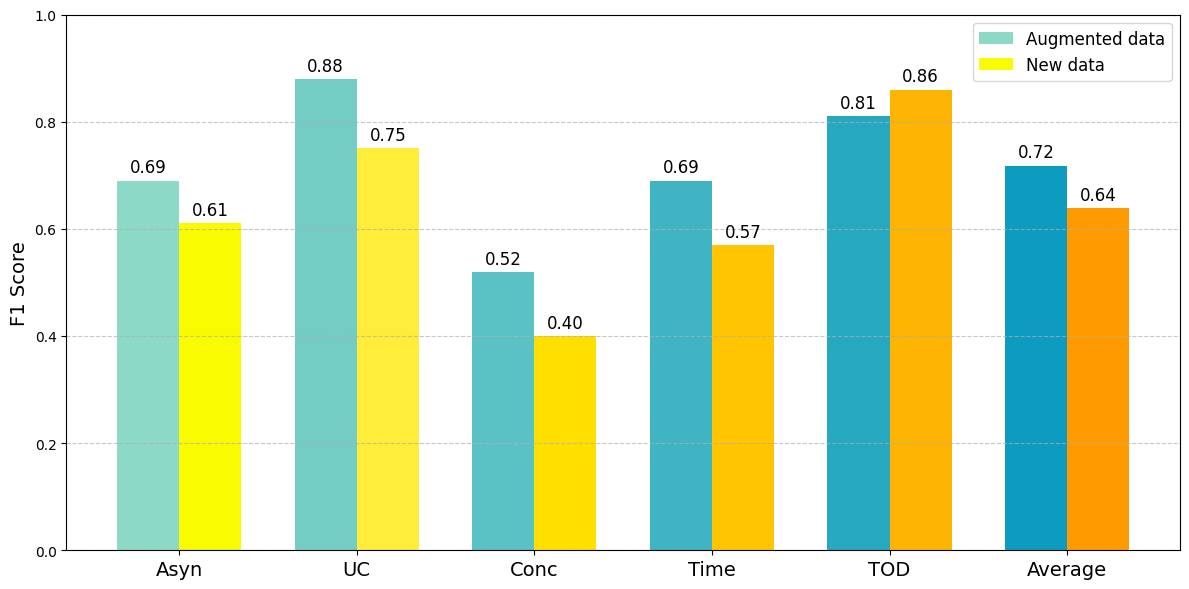}
    \caption{F1 scores by category of testing set~1 vs testing set~2 from Experiment~2}
    \label{fig:base-performance}
\end{figure}

The results in Figure~\ref{fig:base-performance} reveal a systematic bias: the model performs consistently better on augmented versions of tests seen during training, with an average F1 score difference of 8\% compared to new test cases. This performance gap is evident across most flaky test categories, except for Test Order Dependency. The findings suggest that augmentation introduces artifacts that make augmented tests easier to classify, though it also helps the model learn useful patterns for unseen data.

The exception in Test Order Dependency, where the model performs slightly better on new data, highlights potential differences in how augmentation interacts with this category or suggests that test order dependencies have more distinctive patterns that generalize well without relying on augmentation.

These results underscore a key challenge with data augmentation: while it improves overall performance (as shown in Experiment 1), it appears to do so partly by introducing systematic patterns that may not align with real-world flaky test variations. This highlights the need for augmentation techniques that can generate more naturally varied test cases while retaining the essential characteristics of flaky tests.

\begin{table}[t]

\begin{center}
\begin{tabular}{l|c|c|c}
\hline
\textbf{Category} & \textbf{Testing set~1} & \textbf{Testing set~2} & \textbf{Difference} \\
\hline
Async & 61\% & 69\% & +8\% \\
UC & 75\% & 88\% & +13\% \\
Conc & 40\% & 52\% & +12\% \\
Time & 57\% & 69\% & +12\% \\
TOD & 86\% & 81\% & -5\% \\
\hline
\textbf{Average} & \textbf{64\% }& \textbf{72\% }& \textbf{+8\%}\\
\hline
\end{tabular}
\end{center}
\caption{F1 Scores by Category for Testing set~1 and~2 in Experiment~2}
\label{tab:f1_scores}
\end{table}

\subsection{Threats to Validity}

We have intentionally applied our analysis to augmented data created by third-party researchers (Akli et al.~\cite{AHH+23}) to reduce self-confirmatory bias. 

Akli et al. used an adapted SMOTE approach and the results we obtained from evaluating their augmented FlakyCat data set may not generalize to other augmentation strategies or other variations of SMOTE. Furthermore, because our results are based on a single dataset (FlakyCat) they may not generalize to other datasets either. 

In addition to limitations in the generalizability of our results to different augmentation practices and data sets, there are also possible limits to generalizing the results to different sources of bias.

\subsection{Implications for Practice}

Based on our preliminary results, we see several practical implications:
\begin{itemize}
\item The systematic performance gap between augmented and new test cases (8\% average F1 score difference) suggests \textbf{there is a need to maintain a completely separate validation set of original, non-augmented data during model evaluation and testing}. This practice will help provide a more realistic assessment of the model performance rather than an evaluation that overestimates performance due to bias. Furthermore, this also means that data set creators need to clearly identify and label augmented data to support unbiased use by third-party researchers.
\item The varying effectiveness of augmentation across different flaky test categories (e.g., significant improvements for Async Wait versus mixed results for Test Order Dependency) indicates \textbf{there is a potential benefit of category-specific augmentation strategies rather than a one-size-fits-all approach to data augmentation}.

\item The implications of bias in data augmentation likely extend beyond flaky test detection. \textbf{Researchers in areas such as defect prediction and fault localization can benefit from assess bias in augmented data as well as strategies to reduce any identified bias and thus equitably improve model performance across diverse contexts}.
\end{itemize}

%By adopting these practices, practitioners can develop more fair, reliable, and robust software systems.

\section{Conclusions}

% This paper presents a systematic case study of augmentation-induced bias involving a model for flaky test classification. Our experiments showed that there is clear benefit from using the augmented FlakyCat data set and there is also bias introduced when augmented samples are included in the training data as well as the testing data. Specifically, we observed an average F1 score difference of 8\% between the evaluations of the augmented versus truly independent test cases. This bias varied between different flaky test categories, from minimal impact in Test Order Dependency (5\% difference) to more substantive differences in Unordered Collections (13\%). This suggests that in the case of the FlakyCat data set, the effectiveness of augmentation and the potential bias depend heavily on the underlying code patterns.

% Based on these findings, we recommend maintaining strictly separated validation sets of non-augmented data for model evaluation and testing as well as considering category-specific augmentation strategies rather than applying uniform techniques to maximize augmentation benefits. Although data augmentation remains valuable for addressing data scarcity and lack of diversity in software engineering, our study indicates the importance of careful use of augmented data, as well as the need for evaluations that can detect potential augmentation-induced biases and ensure that reported results accurately reflect real-world model performance.

This paper presents a case study on augmentation-induced bias in a flaky test classification model. Our experiments reveal both benefits and biases from using the augmented FlakyCat dataset. Specifically, we observed an average F1 score difference of 8\% between evaluations on augmented and truly independent test cases, with bias varying across flaky test categories, ranging from minimal in Test Order Dependency (5\%) to more substantial in Unordered Collections (13\%). This indicates that augmentation effectiveness and bias depend on underlying code patterns.

Based on our findings, we recommend using strictly separated, non-augmented validation sets for evaluation and adopting category-specific augmentation strategies to maximize benefits. While augmentation is valuable for addressing data scarcity and diversity issues in software engineering, our findings highlight the need for careful use of augmented data and robust evaluations to detect biases and ensure results reflect real-world model performance.

\section{Future Work}

In the future, more studies of augmentation techniques, such as SMOTE and mutation-based methods, are needed to understand not just the benefits of augmentation, but also the risk of bias. %to evaluate their effectiveness across flaky test categories. 
Additionally, a broader investigation into the prevalence, usage practices, and impact of augmented data in real-world software testing practices would provide valuable insights. Lastly, developing best practices for mutation-based augmentation will help refine methodologies and promote their responsible and unbiased application in software engineering.

\bibliographystyle{IEEEtran}
\bibliography{references}

\end{document}